# Maunakea Spectroscopic Explorer (MSE)
# The Prime Focus Subsystems: Requirements and Interfaces

Alexis Hill[1a], Alexandre Blin[b], David Horville[c]
Shan Mignot[c], Kei Szeto[a],

[a] CFHT Corporation, 65-1238 Mamalahoa Hwy, Kamuela, Hawaii 96743, USA
[b] Division Technique de l'INSU, Bureau d'études mécanique, 1 place Aristide Briand, 92195 Meudon, France,
[c] GEPI, Observatoire de Paris, PSL Research University, CNRS, Univ Paris Diderot, Sorbonne Paris Cité, Place Jules Janssen, 92195 Meudon, France

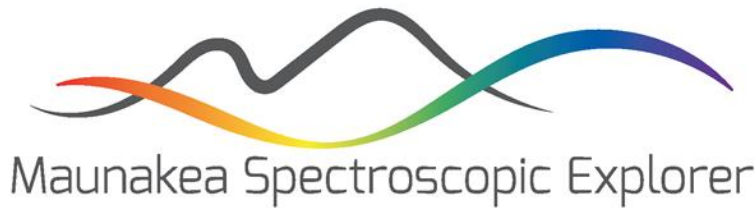

## ABSTRACT

MSE will be a massively multiplexed survey telescope, including a segmented primary mirror which feeds fibers at the prime focus, including an array of approximately four thousand fibers, positioned precisely to feed banks of spectrographs several tens of meters away.

We describe the process of mapping top-level requirements on MSE to technical specifications for subsystems located at the MSE prime focus. This includes the overall top-level requirements based on knowledge of similar systems at other telescopes and how those requirements were converted into specifications so that the subsystems could begin working on their Conceptual Design Phases. We then discuss the verification of the engineering specifications and the compiling of lower-level requirements and specifications into higher level performance budgets (e.g. Image Quality). We also briefly discuss the interface specifications, their effect on the performance of the system and the plan to manage them going forward. We also discuss the opto-mechanical design of the telescope top end assembly and refer readers to more details for instrumentation located at the top end.

**Keywords:** Maunakea Spectroscopic Explorer, multi-object spectrograph, prime focus, requirements, interfaces

## 1. INTRODUCTION

The Maunakea Spectroscopic Explorer (MSE) is an upgrade of the 3.6-m Canada France Hawaii Telescope (CFHT) into a dedicated optical and near-infrared (NIR) spectroscopic survey facility. MSE is planned as 10-meter effective aperture telescope with a 1.5 square degree field of view located at a prime focus. MSE will operate in the optical to near-infrared, at low, moderate and high spectral resolutions. MSE's first-light instrumentation suite takes advantage of this large aperture and field of view by including a massively multiplexed fiber-fed system that will be capable of collecting millions of spectra per year. The multiplexing includes an array of approximately four thousand fibers, positioned precisely to feed banks of spectrographs several tens of meters away.

During the recent Conceptual Design Phase (CoDP) of MSE [1] subsystem designs were developed based on MSE's overall science requirements [2], system architecture and other constraints such as environmental conditions and mass limits. During this process, subsystem designers developed practical designs, based on previous experience with similar hardware. The results of this work were used to inform the plans for the observatory system architecture and operations concept and for creating realistic overall system performance budgets.

---

[1] Email: hill@cfht.hawaii.edu; Telephone: 808-885-3187

With the conceptual designs complete, MSE project office has compared system performance budgets to the science-based requirements. The overall systems engineering process for the project is described in [3] and the process of maximizing sensitivity is described in [4]. MSE is now preparing to proceed with its Preliminary Design Phase (PDP).

This paper discuss the mapping of top level requirements to technical specifications for the subsystems at the prime focus. A description of the prime focus instruments of MSE are included, as well as a more detailed description of the telescope top end assembly conceptual designs. We discuss the requirements, interfaces and constraints which will define the subsystems at the prime focus going forward.

## 2. OBSERVATORY LAYOUT

The overall layout of the observatory is shown in Figure 1. MSE is an altitude-azimuth telescope with a 60-segment primary mirror (M1) and an 11.25-m entrance pupil (10-m effective diameter). M1 reflects light to a wide field corrector/atmospheric dispersion corrector (WFC/ADC) to create the prime focus at the top end of the telescope.

MSE elevation and azimuth structures provide support during observations, over the full range of motion. The elevation structure rotates on small trunnions and supports M1 segments in a mirror cell. At its top end, the elevation structure has six spiders and an interface support ring, which supports all of the prime focus subsystems. The azimuth structure rotates on a large circular track and supports the elevation structure as well as instrument platforms on both sides of the structure. The platforms support a bank of low-moderate resolution spectrographs. A set of high resolution spectrographs are located in the Coudé room of the telescope pier.

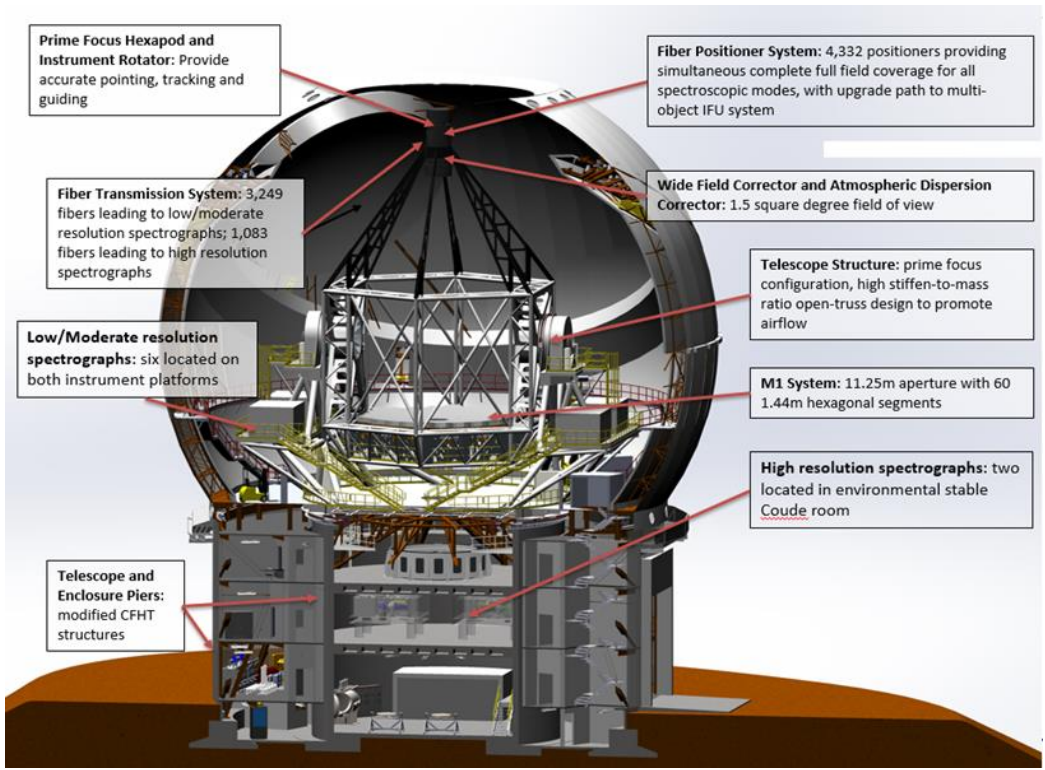

Figure 1. Overall architecture of MSE.

The prime focus is a convex focal surface (see Figure 2) with a 1.52 square degrees field of view (584 mm in diameter). MSE packs 4332 fibers, mounted in fiber positioning actuators, in a hexagonal array inside the field of view, to capture light from individual targets and transmit it to banks of spectrographs. The remaining edges of the field of view are reserved for use by on-axis deployable acquisition and guide cameras (AGC), as well as phasing and alignment camera (PAC) for M1 figuring operations. The array of fibers and the guide cameras are rotated to follow the targets as the telescope system follows observing fields on the sky.

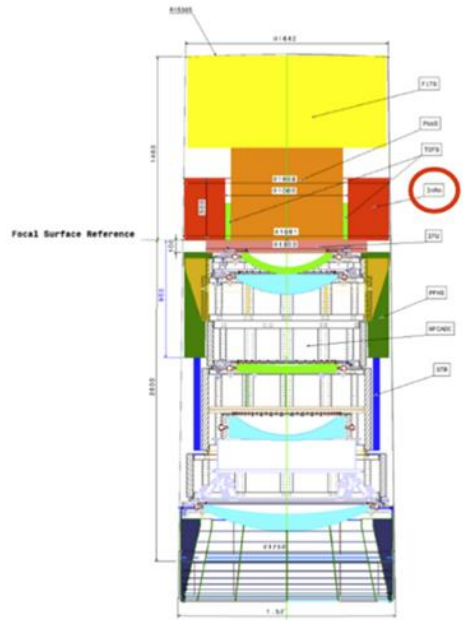

Figure 2. Prime focus subsystems and FOV.

## 2.1 Instrumentation

Science instrumentation located at the prime focus collect and transmit light to the spectrographs.

Fiber Transmission System (FiTS) [5] is a fiber optic relay system, developed at Herzberg Astronomy and Astrophysics (HAA) in Canada, with the primary purpose of transmitting the light from targets in the MSE focal surface to the spectrographs in a highly efficient matter (i.e. with high throughput) over the broad wavelength ranges while being stable, robust and reliable.

Fiber Positioner System (PosS) [6] is an array of identical piezo "tilting spine" actuators, developed by Australian Astronomical Observatory (and named Sphinx), each of which positions the input end of each optical fiber laterally at the focal surface, to ensure efficient light injection into the 4332 fibers. During target acquisition, a closed loop metrology system (FPMS) ensures the fiber inputs are at the desired position on their targets.

The instrumentation suite of MSE, including FiTS and PosS, is discussed in a separate paper [7].

## 2.2 Guiding and Phasing and Alignment Cameras

Camera systems are also located at the prime focus. The Acquisition and Guide System (ACG) is a set of three off-axis cameras with a combined field of view large enough to facilitate guide star acquisition and guiding during observations with MSE. The cameras provide control signals for real-time feedback to telescope guiding, ADC control error and top end (PFHS and InRo) misalignment due to pointing and tracking errors. As well, the cameras are used to develop look-up tables so that adjustments due to gravity and temperature effects can be made.

The Phasing and Alignment camera system (PAC) is planned as a Shack-Hartmann wavefront sensor responsible for the on-axis wavefront quality of MSE. In order to produce wavefronts of acceptable quality, PAC will adjust primary mirror segment pistons and tip/tilts as well as segment surface figuring (via warping harness adjustments).

## 2.3 Top End Assembly

The telescope's Top End Assembly (TEA), developed by DT-INSU in France, is a collection of three related subsystems that are closely integrated with the telescope to support observations. This includes the WFC/ADC, as well as a Prime Focus Hexapod System (PFHS) and Instrument Rotator (InRo).

WFC/ADC is based on the optical design for MSE [8] and provides primary mirror wide field optical aberration correction and atmospheric dispersion correction and delivers the corrected focal surface at the telescope prime focus. DT-INSU designed the WFC barrel, an opto-mechanical assembly, to align, support and protect the optics of this system.

PFHS provides top end subsystem positional correction in five degrees of freedom (focus, decenter and tip/tilt) to compensate for dimensional changes of the telescope structure due to environmental and gravity orientation effects. By making these moves, PFHS maintains the alignment of the WFC barrel to M1, ensures the fibers and the guide cameras are positioned at the focal surface and provides a small offset as part of the ADC control action to allow for atmospheric dispersion correction.

InRo provides field derotation as the telescope follows the sky. All instruments located on the field of view, as well as the telescope's guide cameras, ride on its large-bearing mechanism.

TEA conceptual designs are discussed in more detail later in this paper.

# 3. REQUIREMENTS DEVELOPMENT

MSE strives to take advantage of the excellent site seeing of Maunakea with the best possible sensitivity, massive multiplexing and to be a well-calibrated, survey machine. Sensitivity of MSE is affected by physical effects of individual prime focus subsystems. Sensitivity is quantified in science requirements [2] by the signal to noise ratio (SNR). In MSE, the majority of contributors to SNR are predetermined by the system architecture (e.g. aperture size and field of view). These include quantities such as Noise, Throughput, Injection Efficiency (IE) and Image Quality (IQ). Sensitivity has been maximized in MSE [4], by considering the contributors from the as-designed estimates of subsystem conceptual designs and compiling them in performance budgets.

## 3.1 Throughput

Throughput is the amount of flux from an astronomical target that reaches the detector and is expressed as a percentage of the original target flux. More simply, throughput is the amount of light transmitted by the system. For the prime focus subsystems of MSE, throughput is applies to the WFC/ADC and FiTS (for a summary see Figure 3).

| Wavelength (nm) | 360 | 370 | 400 | 482 | 626 | 767 | 900 | 910 | 950 | 962 | 1235 | 1300 | 1500 | 1662 | 1800 |
|---|---|---|---|---|---|---|---|---|---|---|---|---|---|---|---|
| TEL.WFC/ADC | 54% | 61% | 79% | 87% | 81% | 82% | 84% | 84% | 85% | 85% | 83% | 81% | 73% | 71% | 58% |
| SIP.FiTS (LMR) | 42% | 46% | 58% | 72% | 82% | 85% | 86% | 86% | 84% | 85% | 79% | 79% | 74% | 79% | 61% |
| SIP.FiTS (HR) | 58% | 62% | 70% | 81% | 88% | 89% | 89% | 0% | 0% | 0% | 0% | 0% | 0% | 0% | 0% |

Figure 3. Throughput of prime focus subsystems.

WFC/ADC throughput is dominated by the optical design, especially Fresnel, transmission and absorption losses. WFC/ADC is also limited by available optical blank sizes, so vignetting has an impact on throughput.

For FiTS, throughput effects due to the fibers are considered: Focal Ratio Degradation (FRD), Fresnel and transmission losses. These effects are constraints on the design of the FiTS system but these contributions are not related to the design of the subsystem at the prime focus and are discussed in a separate paper [7].

## 3.2 Noise

Noise is the light that arrives at detectors in the spectrographs from sources other than the object being observed (e.g. sky background and other sources) plus the noise inherent in the detector itself. Noise is not a subject for the prime focus of MSE and is discussed in a separate paper [7].

### 3.3 Image Quality

IQ is a measure of how much an image represented by its point spread function (PSF) is degraded and, for MSE, is expressed as the diametric 80% encircled energy (EE80) of a 2D Moffat distribution on the focal surface at the fibre input. Contributors that affect IQ are grouped: natural site and observatory seeing, M1 segment fabrication and alignment errors, and WFC/ADC fabrication and alignment errors.

Natural site seeing and M1 errors are, of course, out of the control of the prime focus systems of MSE. The subsystems at the prime focus, however, are designed based on the known seeing conditions on Maunakea at the CFHT site.

WFC/ADC optical design focused heavily on IQ performance and includes consideration of errors due to fabrication (i.e. figure errors, glass homogeneity errors) and alignment errors (both during the assembly process and due to flexure during operations) for the WFC/ADC as a rigid body and for individual optical elements.

Observatory seeing can be significantly affected by heat distribution in the dome environment. For that reason, all prime focus components are required to limit or control their heat dissipation, especially near the optical path.

### 3.4 Injection Efficiency

Keeping the fiber input ends aligned with the targets in the sky is critical to the IE of MSE, which is defined as the percentage of flux entering the input fibre with respect to the total flux of a point source at the focal surface. IE is modelled [9] as deviations from the ideal fibre location in two directions, longitudinal (z) and lateral (xy), in units of microns. The maximum amount of attainable flux would enter the fibre at z=0 and xy=0. IE will be degraded by any offset error in the position of the entire fiber array or individual fibers, either laterally or longitudinally. Most of the prime focus subsystems contribute to IE (for a summary, see Figure 4).

| IE variation (um) | Lateral | Longit. | Source |
|---|---|---|---|
| WFC/ADC | 45 max | | Lateral chromatic aberrations |
| | 15 max | | Residual DAR drift after ADC correction |
| InRo | 5 rms | | Control system rotation error (during guiding) |
| | | 30 max | Axis tilt error (during guiding) |
| PFHS | | 5 max | Position errors (during guiding) |
| Guide cameras | 2 rms | | Guiding error |
| PosS | | 30 max | AIV PosS and FiTS |
| | 6 rms | | Closed loop accuracy (after acquisition) |
| | | 80 max | Defocus of spines (but median modelled) |
| | | 1 max | Gravity sag over an observation |
| | 14 max | | Thermal effect over an observation |
| | | 2 max | Thermal effect over an observation |

Figure 4. Injection efficiency of prime focus subsystems.

IE has contributors that include the optical design, Assembly, Integration and Verification (AIV) phase activity and operational errors before and during observations. IE is modelled considering the delivered IQ at the focal surface, as discussed in the previous section. The amount of light that enters the fiber decreases dramatically if the PSF is extended.

The optical design includes contributions inherent to a WFC/ADC that is designed and built within fabrication and alignment tolerances given in the optical design. WFC/ADC has a large contribution to IE, based on two effects. Large displacements are present between any PSFs between wavelengths due to residual chromatic aberrations after atmospheric dispersion correction, even in a theoretically perfect WFC/ADC. The system also will experience atmospheric differential refraction drift during observations over a given zenith range. This is an effect that could be mitigated if it were possible to reposition individual PosS actuators during an observation. This is not currently the baseline for operations, however. Atmospheric dispersion correction is included to reduce the distortion in the system caused by differential atmospheric refraction. This reduces distortion by half over the most of the zenith distance range of motion to provide image quality at the focal surface that enables a small fiber size.

During the set-up for guiding and acquisition, PFHS positions the WFC/ADC plus InRo, PosS, fiber inputs and guide cameras as a rigid body. This compensates for flexure of the telescope structure due to gravity and temperature changes, with a small offset to allow for atmospheric dispersion correction action. This happens once at the beginning of an observation.

As a stand-alone system, PFHS is required to attain and maintain its position to within 0.15 mm laterally and 100 microradians in tip/tilt over the course of any given observation with its own control system. It is not necessary to be more precise than this because the InRo and guide cameras' precise closed loop correction is removes lateral error. Therefore PFHS does not contribute to IE in these four degrees of freedom (lateral and tip/tilt). However neither the telescope pointing nor the InRo are able to correct in the focus direction, so the PFHS is required to position its payload to within +/-0.05 mm and maintain that position over an observation by its own control system.

Errors in the fifth degree of freedom (focus) is not expected to be corrected by PFHS, as a baseline operation.

During observations, InRo will have a 3.5" angular (rotational) accuracy which corresponds to 0.05" on-sky. The angular rate of motion varies as a function of azimuth and elevation of the telescope pointing. Near zenith, there is a 1° diameter "keyhole" in which the rotator is not expected to meet this requirement. As well, the bearing will have an uncorrectable tilt, estimated as 50 urad, corresponding to a maximum defocus at the edges of the field of 30 um, during an exposure. These have been accounted for in all pointing, tracking and guiding error budgets and are acceptable error terms when considering IE. Misalignment of the axis of rotation of the positioners and guide cameras with respect to the InRo rotation axis is not expected to be an issue as long as the update rate of the InRo is fast enough to keep up with the update rate in the guide loop. As well, the guide cameras are in closed loop with the telescope mount control system and InRo will have inherent errors but are not expected to have significant impact on the IE.

During the CoDP phase, InRo positioning requirements were assumed to be limited by bearing tilt and rotational positioning accuracy of the system. It is estimated that there is bearing tilt that will cause about 50 um defocus at the edge of the field over an observation, which will be uncorrectable by the PFHS.

During configuration for an observation, the PosS system acquires its targets, in closed loop with its metrology system. The lateral errors for this are estimated at +/-6 um. While the PosS is in position, the tilted spine will cause defocus errors, depending on how far from vertical the spine has tilted. At its maximum range of motion, the tilt will cause significant defocus errors of +/-80 um. Since the amount of defocus will vary from positioner to positioner, PFHS will position the system to correspond to the mean of the spine tilts. As well, the tilt distribution per field has been modeled [9] for which the defocus is much smaller for the median tilt, so 80 um (max) is a conservative estimate of the error. Regardless, the median distribution was used to estimate IE and the sensitivity requirements are met. There may also be uncorrectable errors (in the PosS) due to gravity sag of the system, temperature changes and so on. These are taken into account in the IE budget as well.

# 4. INTERFACES

The prime focus components of MSE must fit into a small central obscuration (pictured in Figure 5) at the top end of the telescope to minimize the obscured light. Currently, the subsystems are asked to design to fit within that cross-sectional area. The volume has been apportioned into allowable volumes for the subsystems generally and is a challenging constraint. This forms a basis for developing interfaces between the prime focus subsystems that are mechanical in nature.

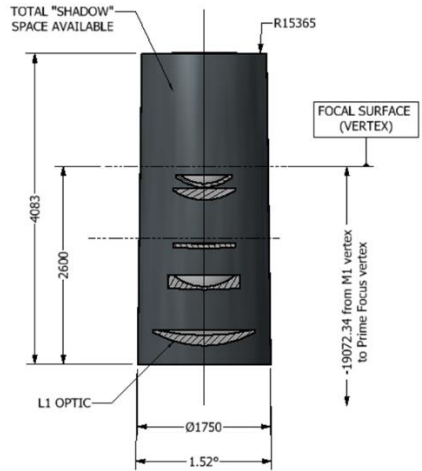

Figure 5. Top end central obscuration.

During CoDP, subsystems were asked to stay within a volume as defined by estimating their needs. Figure 6 shows the arrangement of subdivided space/volume constraints. Not shown are shows the location where two subsystems may reasonably be expected to be mechanically attached.

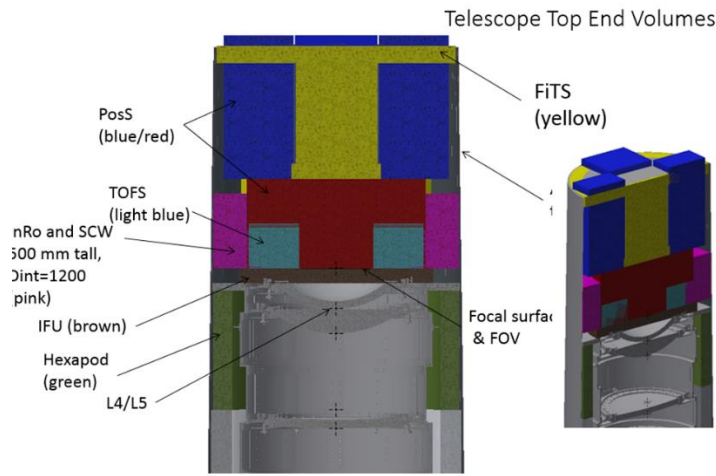

Figure 6. Prime focus interfaces.

Obviously, at this early stage, the interfaces are not well defined. Interfaces will include details such as mechanical attachment, electrical and utilities interfaces, software standards, etc.

In the IE budget, considerations for AIV plan include estimates of errors during the assembly and integration process between interfaces within certain tolerances. This could be the limit of an alignment tolerance, for example, and in the IE budget, these types of considerations apply to specific subsystems or to the MSE Project Office so that control of interfaces and processes remain centralized.

All of the subsystems will be assembled and aligned to each other within estimated tolerances. Particularly, the PosS and FiTS must be integrated with a tolerance on each fiber, individually. When fiber input ends must be aligned to the delivered focal surface to a tight tolerance in the focus direction. This will involve a stack-up of tolerances between PosS and FiTS individual fibers as well as ensuring PosS, PFHS, InRo and WFC/ADC are well aligned. This will require stringent interface definition in the PDP and control throughout all phases.

In the future, interfaces will be developed by a "leading" subsystem (usually the one most impacted by the interface constraints) first and then design iteration between both subsystems will occur. The Project Office will be fully involved in all interface discussions and must approve them as well as any proposed amendments.

## 5. TEA CONCEPTUAL DESIGNS

In this section, we talk about the specific designs for TEA as they are not covered in separate literature elsewhere.

### 5.1 Prime Focus Hexapod System (PFHS)

The hexapod is directly supported by the telescope elevation structure's top end. Its payload includes the WFC/ADC barrel assembly, InRo, PosS, FiTS, ACG and PAC. During the set-up for each observation, the Prime Focus Hexapod System (PFHS) provides a displacement in five degrees of freedom (focus, decenter and tip/tilt), to align the WFC barrel to M1 and ensure PosS and the guide cameras are positioned at the focal surface. Included in this motion is a small offset to allow for the atmospheric dispersion correction action (see WFC/ADC section).

During CoDP, commercial vendors were approached. Symetrie, in France, has proposed a promising solution (Figure 6): a modification of the similar JORAN hexapod, which is similar in size, on the LMT/GTM telescope, Mexico) [10].

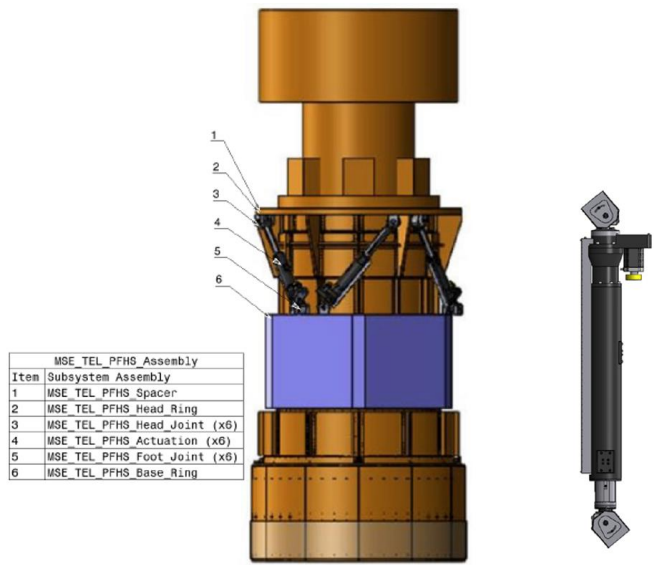

| MSE_TEL_PFHS_Assembly | |
|------|------|
| Item | Subsystem Assembly |
| 1 | MSE_TEL_PFHS_Spacer |
| 2 | MSE_TEL_PFHS_Head_Ring |
| 3 | MSE_TEL_PFHS_Head_Joint (x6) |
| 4 | MSE_TEL_PFHS_Actuation (x6) |
| 5 | MSE_TEL_PFHS_Foot_Joint (x6) |
| 6 | MSE_TEL_PFHS_Base_Ring |

Figure 7. PFHS.

During CoDP, commercial vendors were approached by DT-INSU. Symetrie, in France, has proposed a modification of the similar JORAN hexapod, which is similar in size, on the LMT/GTM telescope . Like other hexapods, this includes six actuators arranged as in the example shown in Figure 6, with a supporting ring at both the interface to the telescope top end and the interface to the payload. In the Joran version of the hexapod, the actuators are brushless motors coupled with a jack. The jack is a ball bearing precision screw with a preloaded nut. Two sensors are included: an absolute sensor with an incremental linear scale and an additional incremental encoder in the motor. This arrangement does not require braking.

Future work will include confirming that cross coupling (a parasitic movement that appears when the trajectory length is very near precision of actuation system) will not dominate the system, particularly ensuring the ±5 um of focus accuracy can be reliably attained. In particular, this could be a limiting factor for providing mid-observation adjustments of the system. As well, it will be confirmed that the Joran hexapod can maintain its positional stability during an observation unless the hexapod is kept under continuous drive loop control, which would likely cause unacceptable high heat dissipation, degrading image quality. However, this concern arose in the context of having a maximum observation of 1 hour and PFHS may be stable (when shut off) for shorter observations. This will be revisited during PDP, including refining the duty cycle of PFHS based on observing fields and as well, the budget for injection efficiency will be reviewed to clarify the requirement.

Under all conditions, PFHS withstands earthquakes without allowing significant damage, especially to items that have a lengthy or expensive recovery, such as the fiber links. This will be confirmed in future work.

PFHS is a well-established technology and will be a reliable and effective component of the overall telescope operation.

## 5.2 Wide Field Corrector/Atmospheric Dispersion Corrector (WFC/ADC)

The Wide Field Corrector/Atmospheric Dispersion Corrector provides wide field optical aberration correction and atmospheric dispersion correction for the 1.5° square field of view and delivers the corrected image at the telescope prime focus. Design of the WFC/ADC included the consideration for IQ but also other factors such as throughput and IE.

The optical design of the WFC/ADC was developed based on many constraints, including Throughput and IQ as discussed. WFC/ADC consists of five lenses (Figure 7), of which three are fused silica and strongly powered (L1, L2, L4), and two are thin lenses of Ohara PBM2Y (L3, L5). AR coating is Solgel coatings which are damaged easily if mishandled. The lenses range in diameter from 1340 mm to 800 mm. Opto-mechanics must be designed to allow for this coating, which involves "spin-coating" these large lenses. Significant vignetting begins to occur at 90% of the field radius.

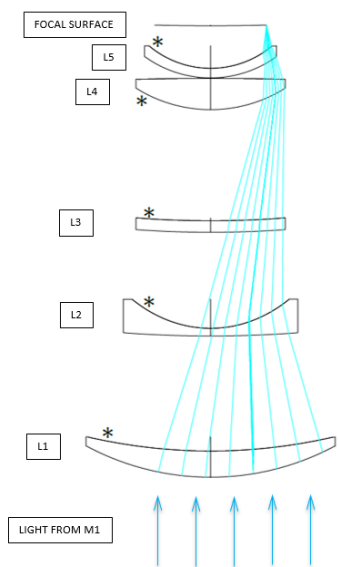

Figure 8. WFC/ADC optics.

DT-INSU has designed an optical barrel to maintain alignment of the optics during science observations. The barrel also protects the optics from the environment. The tolerance analysis in the optical design implied design constraints to maintain the IQ on the optomechanical subsystem of the WFC/ADC. This includes overall tolerances on individual lens elements for decenter between ±0.1 mm to ±0.5 mm, for tip and tilt between ±100 microradians to ±7000 microradians and for defocus about ±0.5 mm in order to meet injection efficiency and image quality requirements. Lenses are mounted in independently adjustable cells to ensure optical alignment is possible at assembly and for handling the optics without damaging the coatings, and then the cells are mounted in a barrel structure. The lens cells allow alignment adjustment in any translation (Tx, Ty, Tz) and for tilts (Rx, Ry). A baffle structure at the entrance to the WFC/ADC prevents stray light from reaching the focal surface, with the whole assembly staying within a maximum central obscuration at the top end.

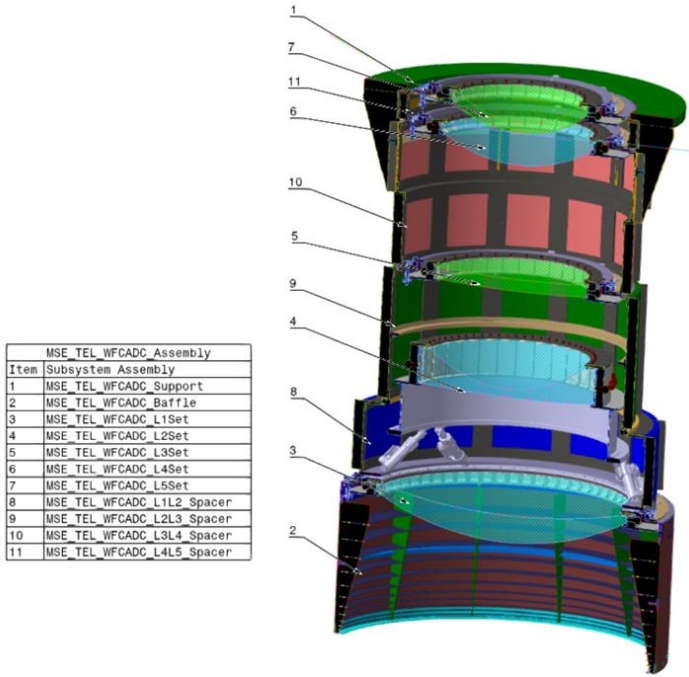

Figure 9. WFC/ADC opto-mechanics.

There is some question about whether the current design as shown will be dimensionally stable over the long term as it is somewhat over-constrained and complicated. This will be explored in future work, with a possible change to a simpler cell and barrel design.

Atmospheric dispersion correction action is accomplished through the combination of three coordinated actions:

- the second lens (L2) of the WFC/ADC is given a lateral shift and tilt,

- the entire WFC assembly is tilted on the Y-Z plane (including L2) as a rigid body using the hexapod mechanism (see PFHS section),

- the whole telescope repoints.

The position of L2 and PFHS are set at their optimal optical location at the beginning of the any given exposure. In Figure 9, the entire WFC/ADC rotates about the point C1 as the global tilt via PFHS while L2 rotates about point C2.

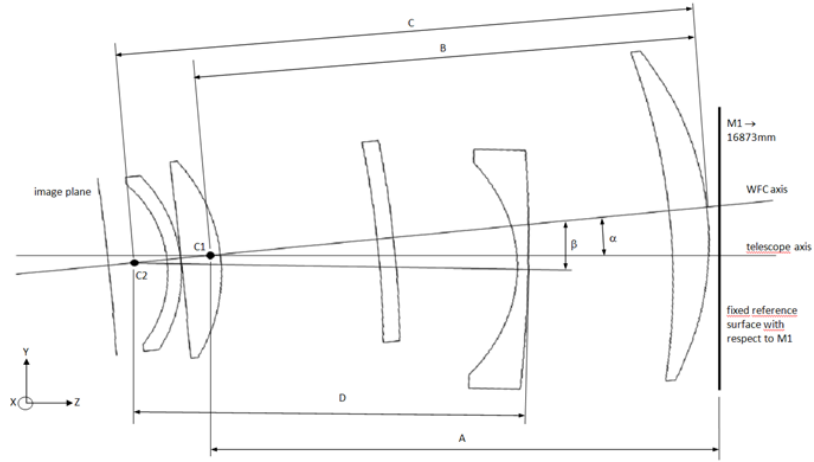

Figure 10. ADC action (L2 and full WFC assembly).

Two mechanisms to provide this ADC action are considered (Figure 11). A hexapod mechanism is proposed to provide the ADC motion, similar to the PFHS. Although this meets requirements and fits in the very tight volume constraints in the area surrounding the L2, it is thought that a more simplified mechanism should be considered, with only two degrees of freedom instead of six. In the figure, each of A, B, C is a cam follower in short tracks with D, E providing tilt motion. This will be explored further in future work.

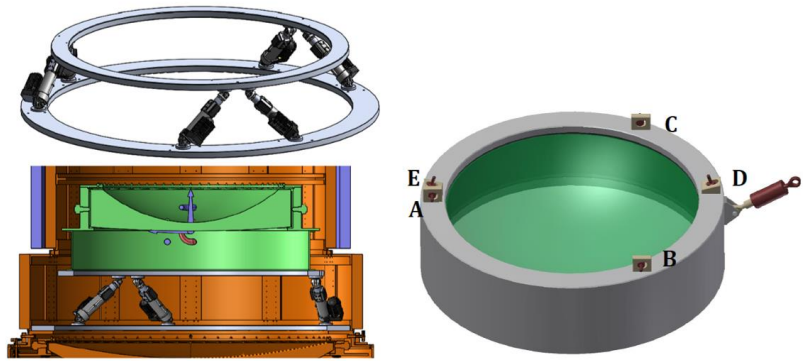

Figure 11. Mechanism to provide ADC (L2 only) action. Hexapod (left) and Cam follower (right).

### 5.3 Instrument Rotator (InRo)

As an alt-azimuth telescope, MSE must derotate the fiber array and guide cameras about the optical axis, with respect to the targets in the field of view. InRo provides this rotation, using a large diameter, precise, rotary bearing system. All instruments located on the field of view, ride on a large-bearing instrument rotator (InRo) mechanism. This rotation is independent of the WFCADC. The guide cameras provide feedback of the angular position and speed with respect to the sky and this is used to ensure the InRo maintains the array of fibers on the sky targets.

InRo also includes a Service Cable Wrap (SCW), for routing needed electrical services (power and data) between systems, as well as utility services (coolant and dry air), between the telescope mount structure and all of the components that need them.

Rotation range is required to be ±180°, to allow for maintenance. During CoDP, this was reduced from ±270°, to allow a more simple design of the SCW in the space available. In practice, the fiber systems, which use the FiTS rotation guide system and not the SCW for fiber management, can likely tolerate a more limited range of motion and will likely be limited to ±90° to avoid damaging fiber optics or affecting their performance by using the full range available. This will be explored in future work as limiting the range of motion has the potential to degrade the observing efficiency of the system.

InRo is be supplied by a commercial vendor. CoDP work by DT-INSU involved speaking to vendors and focused on critical components: the motor and the bearing, to ensure it would be able to meet requirements. DT-INSU approached several commercial vendors in Europe.

A torque motor or gear driven motor (Figure 12) were considered during the CoDP as having the most potential for the size and payload. The torque motor has integrated actuator and encoder and offers high precision, high torque, with no mechanical contact. However, it may have and unacceptable level of heat dissipation near the field of view. This may be mitigated by designing a motor that minimizes dissipation but that may introduce mass and size issues that can affect performance.

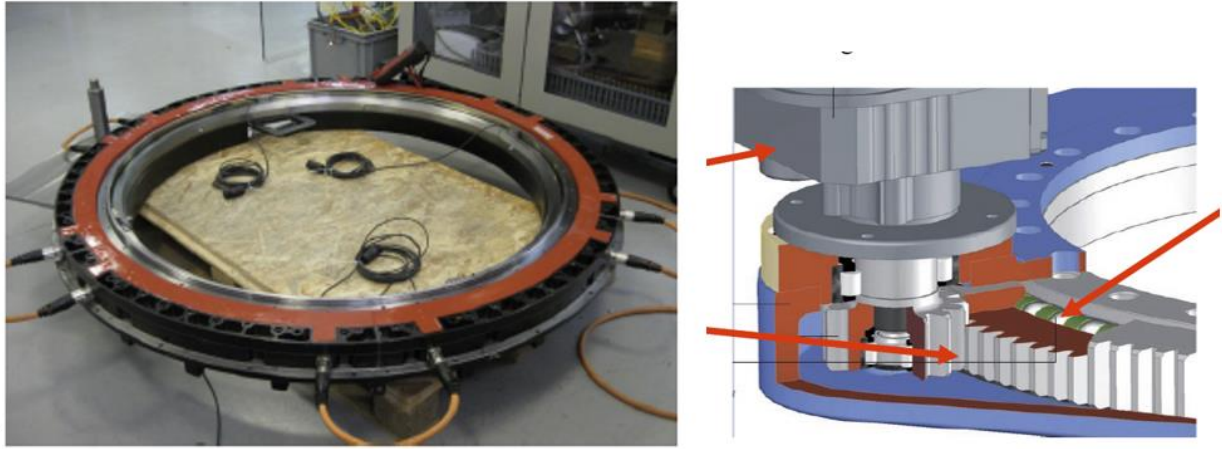

Figure 12. InRo actuator options. Torque motor (left) and Gear slew motor (right).

The gear driven motor includes double roller bearing, motor and encoder and could have two counter-rotating motors to provide both motion and braking. This would likely be a less massive solution but is likely not to provide enough accuracy.

The main constraint on the bearing is to ensure the bearing runout remains as low as possible and to ensure the rotation axis of the rotator is coaxial with the optical axis of the payload. The only appropriate technology identified is a roller bearing. There are 2 types of configuration that will work for InRo: axial-radial cylindrical roller bearings or crossed roller bearing (Figure 13).

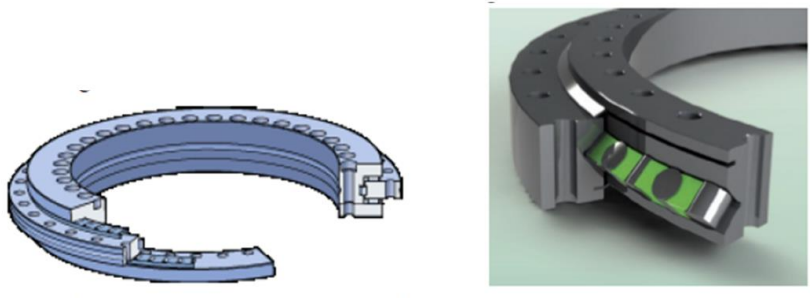

Figure 13. InRo bearing options. Axial-radial roller bearing (left) and cross-roller bearing (right).

Considering these technologies, InRo will have a 3.5" angular (rotational) accuracy which corresponds to 0.05" on-sky. The angular rate of motion varies as a function of azimuth and elevation of the telescope pointing. Near zenith, there is a 1° diameter "keyhole" in which the rotator is not expected to meet this requirement. As well, the bearing will have an uncorrectable tilt, estimated as 50 urad, corresponding to a maximum defocus at the edges of the field of 30 um, during an exposure. These have been accounted for in all pointing, tracking and guiding error budgets and are acceptable error terms when considering injection efficiency.

Similar to PFHS, InRo supports its payload (PoSS and AGC), both during normal observing but also during daytime operations and maintenance, where it supports the payload while cantilevered. Under all conditions, InRo withstands earthquakes without allowing significant damage, especially to items that have a lengthy or expensive recovery, such as the fiber links. This will be confirmed in future work.

Several projects are working with similar sized systems and commercial vendors will be able to design a rotator which meet the size and mass requirements without difficulty. Acceptable mechanisms are readily available for InRo's critical components. Future work will include making decisions on which motor and bearing configuration to choose and developing the design to include the SCW and other interface and support structures. InRo is expected to be a reliable and effective component of the overall telescope operation.

# 6. CONCLUSION

During the conceptual design phase, international partners created conceptual designs of prime focus subsystems based on some general system constraints. Risks and challenging constraints were identified. With the designs in hand, MSE compiled the contributors to the system budgets for Throughput, Injection Efficiency and Image Quality, which are ultimately the quantities that are used to assess whether MSE will meet science requirements.

Given a technical approach that maximizes the use of existing, reliable technology, MSE will be a practical, robust and achievable robust system, to allow it to be a massively multiplexed and powerful survey instrument.

# ACKNOWLEDGEMENTS

The Maunakea Spectroscopic Explorer (MSE) conceptual design phase was conducted by the MSE Project Office, which is hosted by the Canada-France-Hawaii Telescope (CFHT). MSE partner organizations in Canada, France, Hawaii, Australia, China, India, and Spain all contributed to the conceptual design. The authors and the MSE collaboration recognize the cultural importance of the summit of Maunakea to a broad cross section of the Native Hawaiian community.